\pdfoutput=1
\documentclass[11pt]{article}

\usepackage{acl}

\usepackage{times}
\usepackage{latexsym}

\usepackage[T1]{fontenc}
\usepackage[utf8]{inputenc}

\usepackage{microtype}

\usepackage{amsmath}
\usepackage{bm}
\usepackage{amsfonts}
\usepackage{graphicx}
\usepackage{booktabs}
\usepackage{multirow}
\usepackage{float}
\usepackage{makecell}
\usepackage{placeins}

\newcommand*\diff{\mathop{}\!\mathrm{d}}

\newcommand\E{\mathbb{E}}
\newcommand{\Var}{\mathrm{Var}}
\newcommand{\R}{\mathbb{R}}
\newcommand{\Loss}{\mathcal{L}}
\newcommand{\Lnll}{\Loss^{\scalebox{.7}{{NLL}}}}

\newcommand{\x}{\mathbf{x}}
\newcommand{\p}{\mathrm{p}}
\newcommand{\barLnll}{\bar{\Loss}^{\scalebox{.7}{{NLL}}}}

\title{Estimating the Uncertainty in Emotion Attributes \\using Deep Evidential Regression}

\author{Wen Wu$^1$, Chao Zhang$^2$, Philip C. Woodland$^1$\\
$^1$ Department of Engineering, University of Cambridge, Cambridge, UK \\ 
$^2$ Department of Electrical Engineering, Tsinghua University, Beijing, China \\
\texttt{\{ww368, pcw\}@eng.cam.ac.uk; cz277@tsinghua.edu.cn}}
\begin{document}
\maketitle
\begin{abstract}
In automatic emotion recognition (AER), labels assigned by different human annotators to the same utterance are often inconsistent due to the inherent complexity of emotion and the subjectivity of perception. Though deterministic labels generated by averaging or voting are often used as the ground truth, it ignores the intrinsic uncertainty revealed by the inconsistent labels. This paper proposes a Bayesian approach, deep evidential emotion regression (DEER), to estimate the uncertainty in emotion attributes. Treating the emotion attribute labels of an utterance as samples drawn from an unknown Gaussian distribution, DEER places an utterance-specific normal-inverse gamma prior over the Gaussian likelihood and predicts its hyper-parameters using a deep neural network model. It enables a joint estimation of emotion attributes along with the aleatoric and epistemic uncertainties. 
AER experiments on the widely used MSP-Podcast and IEMOCAP datasets showed DEER produced state-of-the-art results for both the mean values and the distribution of emotion attributes\footnote{Code available: https://github.com/W-Wu/DEER}.

\end{abstract}

\section{Introduction}
Automatic emotion recognition (AER) is the task that enables computers to predict human emotional states based on multimodal signals, such as audio, video and text. An emotional state is defined based on either categorical or dimensional theory.
The categorical theory claims the existence of a small number of basic discrete emotions ( {i.e.} anger and happy) that are inherent in our brain and universally recognised~\cite{gunes2011emotion,plutchik2001nature}. Dimensional emotion theory characterises emotional states by a small number of roughly orthogonal fundamental continuous-valued bipolar dimensions~\cite{schlosberg1954three,nicolaou2011continuous} such as {valence-arousal} and {approach–avoidance}~\cite{russell1977evidence,russell1980circumplex,grimm2007primitives}. These dimensions are also known as {emotion attributes}, which allow us to model more subtle and complex emotions and are thus more common in psychological studies. As a result, AER includes a classification approach based on emotion-class-based labels and a regression approach based on attribute-based labels. This paper focuses on attribute-based AER with speech input.

Emotion annotation is challenging due to the inherent ambiguity of mixed emotion, the personal variations in emotion expression, the subjectivity in emotion perception,  {etc.} Most AER datasets use multiple human annotators to label each utterance, which often results in inconsistent labels, either as emotion categories or attributes. This is also a typical manifestation of the intrinsic data uncertainty, also referred to as {aleatoric uncertainty}~\cite{matthies2007quantifying,der2009aleatory}, that arises from the natural complexity of emotion data.
It is common to replace such inconsistent labels with deterministic labels obtained by majority voting~\cite{Busso2008IEMOCAP,MSP-IMPROV} or (weighted) averages \cite{RECOLA,MSP-podcast,kossaifi2019sewa,grimm2005evaluation}. However, this causes a loss of data samples when a majority agreed emotion class doesn't exist~\cite{Majumder2018, Poria2018,wu2021emotion} and also ignores the discrepancies between annotators and the aleatoric uncertainty in emotion data.

In this paper, we propose to model the uncertainty in emotion attributes with a Bayesian approach based on deep evidential regression~\cite{amini2020deep}, denoted deep evidential emotion regression (DEER).
In DEER, the inconsistent human labels of each utterance are considered as observations drawn independently from an unknown Gaussian distribution. 
To probabilistically estimate the mean and variance of the Gaussian distribution, a normal inverse-gamma (NIG) prior is introduced, which places a Gaussian prior over the mean and an inverse-gamma prior over the variance. 
The AER system is trained to predict the hyper-parameters of the
NIG prior for each utterance by maximising the per-observation-based marginal likelihood of each observed label under this prior. 
As a result, DEER not only models the distribution of emotion attributes but also learns both the aleatoric uncertainty and the {epistemic uncertainty}~\cite{der2009aleatory} without repeating the inference procedure for sampling. {Epistemic uncertainty}, also known as model uncertainty, is associated with uncertainty in model parameters that best explain the observed data. 
Aleatoric and epistemic uncertainty are combined to induce the {total uncertainty}, also called predictive uncertainty, that measures the confidence of attribute predictions.
As a further improvement, a novel regulariser is proposed based on the mean and variance of the observed labels to better calibrate the uncertainty estimation. 
The proposed methods were evaluated on the MSP-Podcast and IEMOCAP datasets. 

The rest of the paper is organised as follows. Section~\ref{sec: related work} summarises related work. Section~\ref{sec: method} introduces the proposed DEER approach. Sections~\ref{sec: exp setup} and \ref{sec: exp} present the experimental setup and results respectively, followed by the conclusion.

\section{Related Work}
\label{sec: related work}
There has been previous work by AER researchers to address the issue of inconsistent labels. For emotion categories, a single ground-truth label can be obtained as either a continuous-valued mean vector representing emotion intensities~\cite{Fayek_2016,Ando_2018}, or as a multi-hot vector obtained based on the existence of emotions~\cite{zhang2020multi,ju2020transformer}. Recently, distribution-based approaches have been proposed, which consider the labels as samples drawn from emotion distributions ~\cite{chou2022exploiting,wu2022estimating}.

For emotion attributes, annotators often assign different values to the same attribute of each utterance. \citet{davani2022dealing} proposed a multi-annotator model which contains multiple heads to predict each annotator's judgement. This approach is computationally viable only when the number of annotators is relatively small. The method requires sufficient annotations from each annotator to be effective. \citet{deng2012confidence} derived confidence measures based on annotator agreement to build emotion-scoring models.
\citet{han2017hard,han2021exploring} proposed predicting the standard deviation of the attribute label values as an extra task in the multi-task training framework. 
\citet{dang2017investigation,dang2018dynamic} included annotator variability as a representation of uncertainty in a Gaussian mixture regression model.
These techniques take the variance of
human annotations either as an extra target or as an extra input. 
More recently, Bayesian deep learning has been introduced to the task, which models the uncertainty in emotion annotation without explicitly using the variance of human annotations. These include the use of Gaussian processes~\cite{atcheson2018demonstrating,atcheson2019using}, variational auto-encoders~\cite{sridhar2021generative}, Bayesian neural networks~\cite{prabhu2021end}, Monte-Carlo 
dropout~\cite{sridhar2020modeling} and sequential Monte-Carlo methods~\cite{markov2015dynamic,wu2022novel}.

So far, these methods have not distinguished aleatoric uncertainty from epistemic uncertainty which are defined in the introduction. 
Our proposed DEER approach can simultaneously model these two uncertainties.
In addition, our approach is more generic. It has no limits on the number of annotators, the number of annotators per utterance, and the number of annotations per annotator, and thus can cope with large crowd-sourced datasets. 

\section{Deep Evidential Emotion Regression}
\label{sec: method}
\subsection{Problem setup}
In contrast to Bayesian neural networks that place priors on model parameters~\cite{blundell2015weight,kendall2017uncertainties}, evidential deep learning~\cite{sensoy2018evidential, malinin2018predictive,amini2020deep} places priors over the likelihood function. Every training sample adds support to a learned higher-order prior distribution called the {evidential distribution}. Sampling from this distribution gives instances of lower-order likelihood functions from which the data was drawn.

Consider an input utterance $\x$ with $M$ emotion attribute labels $y^{(1)},\ldots,y^{(M)}$ provided by multiple annotators. Assuming $y^{(1)},\ldots,y^{(M)}$ are observations drawn  {i.i.d.} from a Gaussian distribution with unknown mean $\mu$ and unknown variance $\sigma^2$, where $\mu$ is drawn from a Gaussian prior and $\sigma^2$ is drawn from an inverse-gamma prior:
\begin{equation}
   \nonumber
    y^{(1)},\ldots, y^{(M)} \sim \mathcal{N}(\mu, \sigma^2)
\end{equation}
\begin{equation}
   \nonumber
    \mu \sim \mathcal{N}(\gamma, \sigma^2 \upsilon^{-1}), \quad \sigma^2 \sim \Gamma^{-1}(\alpha, \beta)
\end{equation}
where $\gamma\in\mathbb{R}$, $\upsilon>0$, and $\Gamma(\cdot)$ is the gamma function with $\alpha>1$ and $\beta>0$.

Denote $\{\mu,\sigma^2\}$ and $\{\gamma,\upsilon,\alpha,\beta\}$ as $\boldsymbol{\Psi}$ and $\boldsymbol{\Omega}$. The posterior $\p(\boldsymbol{\Psi}|\boldsymbol{\Omega})$ is a NIG distribution, which is the Gaussian conjugate prior:
\begin{equation}
 \begin{aligned}
    \nonumber \p(\boldsymbol{\Psi}|\boldsymbol{\Omega})&=\p(\mu|\sigma^2,\boldsymbol{\Omega}) \, \p(\sigma^2|\boldsymbol{\Omega})\\
    &= \mathcal{N}(\gamma, \sigma^2 \upsilon^{-1}) \, \Gamma^{-1}(\alpha, \beta)\\
    &= \frac{\beta^{\alpha}\sqrt{\upsilon}}{\Gamma(\alpha)\sqrt{2 \pi \sigma^{2}}} \left(\frac{1}{\sigma^{2}}\right)^{\alpha+1}\\
    &\cdot \exp \left\{-\frac{2 \beta+\upsilon(\gamma-\mu)^{2}}{2 \sigma^{2}}\right\}
\end{aligned}   
\end{equation}
Drawing a sample $\boldsymbol{\Psi}_i$ from the NIG distribution yields a single instance of the likelihood function $\mathcal{N}(\mu_i, \sigma^2_i)$. The NIG distribution therefore serves as the {higher-order}, {evidential} distribution on top of the unknown {lower-order} likelihood distribution from which the observations are drawn. The NIG hyper-parameters $\boldsymbol{\Omega}$ determine not only the location but also the uncertainty 
associated with the inferred likelihood function. 

By training a deep neural network model to output the hyper-parameters of the evidential distribution, evidential deep learning allows the uncertainties to be found
by analytic computation of the maximum likelihood Gaussian without the need for repeated inference for sampling~\cite{amini2020deep}. Furthermore, it also allows an effective estimate of the aleatoric  uncertainty computed as the expectation of the variance of the Gaussian distribution, 
as well as the epistemic uncertainty defined as the variance of the predicted Gaussian  mean.
Given an NIG distribution, the prediction, aleatoric, and epistemic uncertainty can be computed as:
\begin{align}
    \nonumber&\text{Prediction:} \, \E[\mu] = \gamma\\
    \nonumber&\text{Aleatoric:} \, \E[\sigma^2] = \frac{\beta}{\alpha-1}, \quad \forall\, \alpha > 1\\
    \nonumber&\text{Epistemic:} \,  \Var[\mu] = \frac{\beta}{\upsilon(\alpha-1)}, \quad \forall\, \alpha > 1
\end{align}

\subsection{Training}
The training of DEER is structured as fitting the model to the data while enforcing the prior to calibrate the uncertainty when the prediction is wrong. 

\subsubsection{Maximising the data fit} 
\label{sec: nll}
The likelihood of an observation $y$ given the evidential distribution hyper-parameters $\boldsymbol{\Omega}$ is computed by marginalising over the likelihood parameters $\boldsymbol{\Psi}$: 
\begin{equation}
\begin{aligned}
\p(y | \boldsymbol{\Omega}) &= \int_{\boldsymbol{\Psi}} \p(y|\boldsymbol{\Psi}) \p(\boldsymbol{\Psi} | \boldsymbol{\Omega}) \, \diff\boldsymbol{\Psi}\\
&=\mathbb{E}_{\p(\boldsymbol{\Psi} | \boldsymbol{\Omega})}[\p(y|\boldsymbol{\Psi})]\label{eqn: marginal}
\end{aligned}
\end{equation}
An analytical solution exists in the case of placing an NIG prior on the Gaussian likelihood function: 
\begin{align}
\p(y|\boldsymbol{\Omega}) 
\nonumber&= \frac{\Gamma(1/2 + \alpha)}{\Gamma(\alpha) }  \sqrt{\frac{\upsilon}{\pi}}
\left(2\beta (1+\upsilon)\right)^\alpha\\
\nonumber&\cdot \left( \upsilon (y-\gamma)^2 + 2\beta (1+\upsilon)\right)^{-(\frac{1}{2}+\alpha)}\\
&= \text{St}_{2\alpha}\left(y| \gamma, \frac{\beta(1+ \upsilon)}{\upsilon\,\alpha}\right)
\label{eq:post_pred}
\end{align}
where $\text{St}_{\nu}\left(t|r,s\right)$ is the {Student's t-distribution} evaluated at $t$ with location parameter $r$, scale parameter $s$, and $\nu$ degrees of freedom. The predicted mean and  variance can be computed analytically as
\begin{equation}
  \begin{aligned}
    \E[y]=\gamma, \quad \Var[y] = \frac{\beta (1+\upsilon) }{\upsilon (\alpha -1)}
    \label{eqn: predicted mean var}
\end{aligned}  
\end{equation}
$\Var[y]$ represents  the total uncertainty of model prediction, which is equal to the 
summation of the aleatoric uncertainty $\E[\sigma^2]$ and epistemic uncertainty $\Var[\mu]$ according to {the law of total variance}:
\begin{align*}
    \Var[y] &= \E[\Var[y|\Psi]] + \Var[\E[y|\Psi]]\\
    &= \E[\sigma^2] + \Var[\mu]
\end{align*}

To fit the NIG distribution, the model is trained by maximising the sum of the marginal likelihoods of each human label $y^{(m)}$. %For input $\x$ with labels $y^{(1)},\ldots, y^{(M)}$  from $M$ annotators, 
The negative log likelihood (NLL) loss can be computed as
\begin{align}
    % \begin{aligned}
&\Lnll(\mathbf{\Theta})= - \frac{1}{M} \sum_{m=1}^M \log \p(y^{(m)}|\boldsymbol{\Omega}) \label{eqn: nll-per-sample}\\
&= -\frac{1}{M}  \sum_{m=1}^M \log \left[ \text{St}_{2\alpha}\left(y^{(m)}| \gamma, \frac{\beta(1+ \upsilon)}{\upsilon\,\alpha} \right) \right] \nonumber
\end{align}
This is our proposed {per-observation-based NLL loss}, which takes each observed label into consideration for AER. This loss serves as the first part of the objective function for training a deep neural network model $\mathbf{\Theta}$ to predict the  hyper-parameters $\{\gamma,\upsilon,\alpha,\beta\}$ to fit all observed labels of $\mathbf{x}$.

\subsubsection{Calibrating the uncertainty on errors}
The second part of the objective function regularises training by calibrating the uncertainty based on the incorrect predictions. A novel regulariser is formulated which contains two terms: $\mathcal{L}^{\mu}$ and $\Loss^\sigma$ that respectively regularises the errors on the estimation of the mean $\mu$ and the variance $\sigma^2$ of the Gaussian likelihood.

The first term $\mathcal{L}^{\mu}$ is proportional to the error between the model prediction and the average of the observations:
\begin{align}
\Loss^\mu(\mathbf{\Theta}) = \Phi\,|\Bar{y} - \mathbb{E}[\mu]| \nonumber
\end{align}
where $|\cdot|$ is L1 norm,  $\bar{y}=\frac{1}{M} \sum\nolimits_{m=1}^M y^{(m)}$ is the averaged label which is usually used as the ground truth in regression-based AER, and $\Phi$ is an uncertainty measure associated with the inferred posterior.
The reciprocal of the total uncertainty is used as $\Phi$ in this paper, which can be calculated as
\begin{align}
    \Phi = \frac{1}{\Var[y]}  = \frac{\upsilon (\alpha -1)}{\beta (1+\upsilon)} \nonumber
\end{align}
The regulariser imposes a penalty when there's an error in prediction and dynamically scales it by dividing by the total uncertainty of inferred posterior. 
It penalises the cases where the model produces an incorrect prediction with a small uncertainty, thus preventing the model from being over-confident. For instance, if the model produces an error with a small predicted variance, $\Phi$ is large, resulting in a large penalty.
Minimising the regularisation term enforces the model to produce accurate prediction or increase uncertainty when the error is large.

In addition to imposing a penalty on the mean prediction as in \citet{amini2020deep}, a second term $\Loss^\sigma$ is proposed in order to calibrate the estimation of the aleatoric uncertainty. As discussed in the introduction, aleatoric uncertainty in AER is shown by the different emotional labels given to the same utterance by different human annotators. This paper uses the variance of the observations to describe the aleatoric uncertainty in the emotion data. The second regularising term is defined as:
\begin{align}
    \Loss^\sigma(\mathbf{\Theta}) 
&= \Phi\,|\bar{\sigma}^2  - \mathbb{E}[\sigma^2]| \nonumber
\end{align}
where $\bar{\sigma}^2= \frac{1}  {M}\sum_{m=1}^M(y^{(m)} - \Bar{y})^2$.

\subsection{Summary and implementation details}  
For an AER task that consists of $N$ emotion attributes, DEER trains a deep neural network model to simultaneously predict the hyper-parameters $\{\mathbf{\Omega}_1,\ldots,\mathbf{\Omega}_N\}$ associated with the $N$ attribute-specific NIG distributions, where $\mathbf{\Omega}_n=\{\gamma_n,\upsilon_n,\alpha_n,\beta_n\}$. A DEER model thus has $4N$ output units. 
The system is trained by minimising the total loss  {w.r.t.} $\mathbf{\Theta}$ as:
\begin{align}
\Loss_\text{total}(\bm{\Theta}) &= \sum_{n=1}^N \epsilon_n \Loss_n(\bm{\Theta}) \label{eqn: MTL loss}\\
\Loss_n(\bm{\Theta}) &= \Lnll_n(\bm{\Theta}) \nonumber\\
&+ \lambda_n \left[\Loss_n^\mu(\bm{\Theta}) + \Loss_n^\sigma(\bm{\Theta}) \right] \label{eqn: proposed edl loss}
\end{align}
where  $\epsilon_n$ is the weight satisfying $\sum_{n=1}^N \epsilon_n = 1$, $\lambda_n$ is the scale coefficient that trades off the training between data fit and uncertainty regulation.

At test-time, the predictive posteriors are $N$ separate Student's t-distributions $\p(y|\boldsymbol{\Omega}_1)$$,\p(y|\boldsymbol{\Omega}_2)$ $,\ldots,$ $\p(y|\boldsymbol{\Omega}_N)$, each of the same form as derived in Eqn.~\eqref{eq:post_pred}\footnote{Since NIG is the Gaussian conjugate prior, the posterior is in the same parametric family as the prior. Therefore, the predictive posterior has the same form as the marginal likelihood. Detailed derivations see Appendix~\ref{sec:appendix}.}.
Apart from obtaining a distribution over the emotion attribute of the speaker, DEER also allows analytic computation of the uncertainty terms, as summarised in Table~\ref{tab: smry formula}.
\begin{table}[H]
    \centering
    \scalebox{0.88}{
    \begin{tabular}{cc}
    \toprule
    \textbf{Term} & \textbf{Expression}\\
    \midrule
     Predicted mean &  $\E[y]=\E[\mu]=\gamma$ 
    \\
    \midrule
    \makecell{Predicted variance\\(Total uncertainty)} & $\Var[y]=\frac{\beta (1+\upsilon) }{\upsilon (\alpha -1)}$ \\
    \midrule
    Aleatoric uncertainty&
    $\E[\sigma^2]=\frac{\beta}{\alpha-1}$ \\
    \midrule
    Epistemic uncertainty 
    & $\Var[\mu] = \frac{\beta}{\upsilon(\alpha-1)}$ \\ 
    \bottomrule
    \end{tabular}}
    \caption{Summary of the uncertainty terms.}
    \label{tab: smry formula}
\end{table}

\section{Experimental Setup}
\label{sec: exp setup}
\subsection{Dataset}
The MSP-Podcast~\cite{MSP-podcast} and IEMOCAP datasets~\cite{Busso2008IEMOCAP} were used in this paper. The annotations of both datasets use $N=3$ with valence, arousal (also called activation), and dominance as the emotion attributes.
MSP-Podcast contains natural English speech from podcast recordings and is one of the largest publicly available datasets in speech emotion recognition. A seven-point Likert scale was used to evaluate valence (1-negative vs 7-positive), arousal (1-calm vs 7-active), and dominance (1-weak vs 7-strong). The corpus was annotated using crowd-sourcing. Each utterance was labelled by at least 5 human annotators and has an average of 6.7 annotations per utterance. Ground-truth labels were defined by the average value. Release 1.8 was used in the experiments, which contains 73,042 utterances from 1,285 speakers amounting to more than 110 hours of speech. The average variance of the labels assigned to each sentence is 0.975, 1.122, 0.889 for valence, arousal, and dominance respectively. The standard splits for training (44,879 segments), validation (7,800 segments) and testing (15,326 segments) were used in the experiments.

The IEMOCAP corpus is one of the most widely used AER datasets. It consists of approximately 12 hours of English speech including 5 dyadic conversational sessions performed by 10 professional actors with a session being a conversation between two speakers. There are in total 151 dialogues including 10,039 utterances. Each utterance was annotated by three human annotators using a five-point Likert scale.  Again, ground-truth labels were determined by taking the average. The average variance of the labels assigned to each sentence is 0.130, 0.225, 0.300 for valence, arousal, and dominance respectively. Unless otherwise mentioned, systems on IEMOCAP were evaluated by training on Session 1-4 and testing on Session 5.

\subsection{Model structure}
\begin{figure}[tb]
    \centering
    \includegraphics[width=0.9\linewidth]{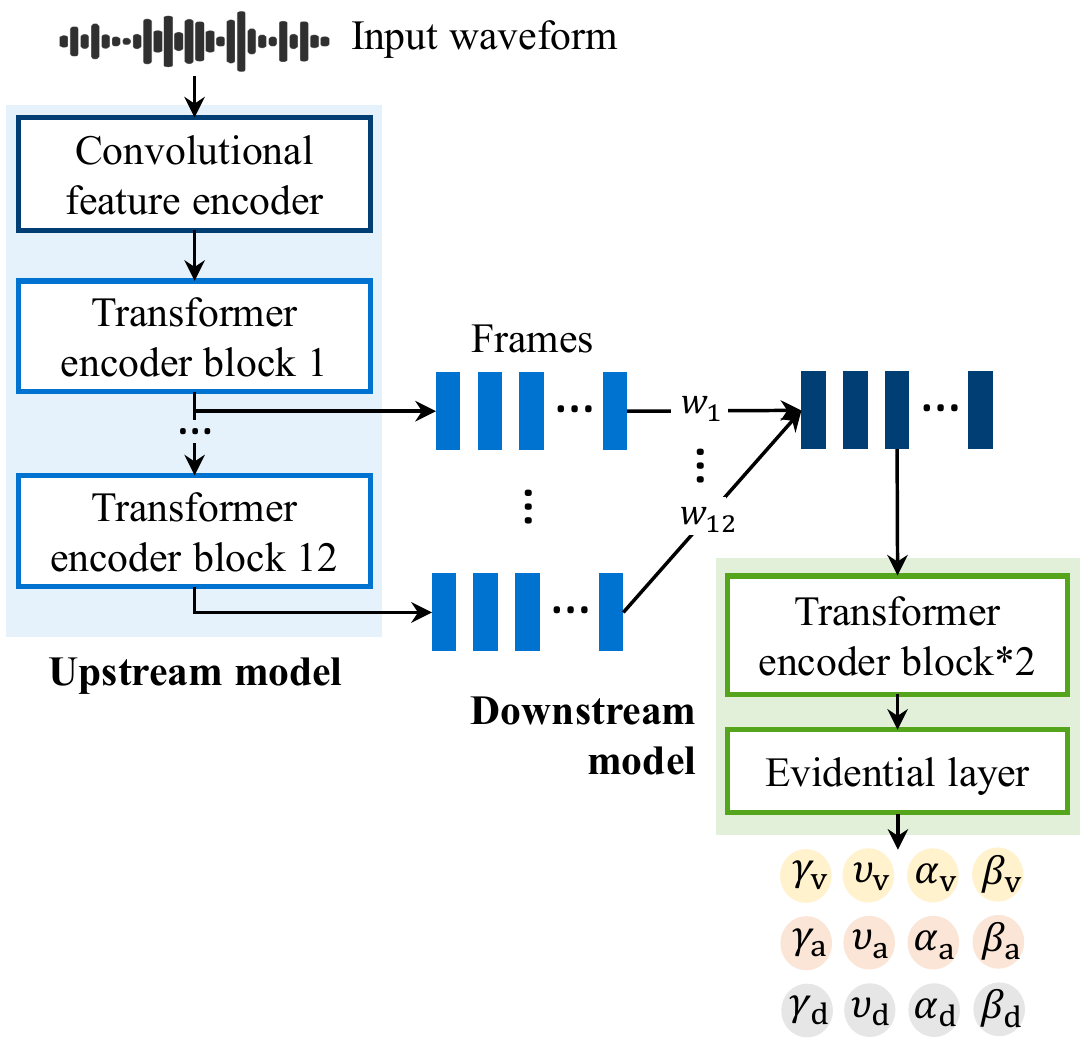}
    \caption{Illustration of the model structure. Weights $w_1, \ldots, w_{12}$ for the weighted sum of the 12 Transformer encoder outputs are trainable  and satisfy $\sum_{i=1}^{12} w_i =1$.}
    \label{fig: struc}
\end{figure}

The model structure used in this paper follows the upstream-downstream framework~\cite{yang2021superb}, as illustrated in Figure~\ref{fig: struc}. WavLM~\cite{chen2022wavlm} was used as the upstream model, which is a speech foundation model pre-trained by self-supervised learning. The BASE+ version\footnote{https://huggingface.co/microsoft/wavlm-base-plus} of the model was used in this paper
which has 12 Transformer encoder blocks with 768-dimensional hidden states and 8 attention heads. 
The parameters of the pre-trained model were frozen and the weighted sum of the outputs of the 12 Transformer encoder blocks was used as the speech embeddings and fed into the downstream model.

The downstream model consists of two 128-dimensional Transformer encoder blocks with 4-head self-attention, followed by an evidential layer that contains four output units for each  of the three attributes, which has a total of 12 output units.
The model contains 0.3M trainable parameters. 
A Softplus activaton\footnote{$\operatorname{Softplus}(x)=\ln (1+\exp(x))$} was applied to $\{\upsilon, \alpha, \beta\}$ to ensure $\upsilon, \alpha, \beta >0$ with an additional +1 added to $\alpha$ to ensure $\alpha > 1$. A linear activation was used for $\gamma \in \R$.  The proposed DEER model was trained to simultaneously learn three evidential distributions for the three attributes. The weights in Eqn.~\eqref{eqn: MTL loss} were set as $\epsilon_v = \epsilon_a = \epsilon_d = {1}/{3}$. The scale coefficients were set to $\lambda_v=\lambda_a=\lambda_d=0.1$ for Eqn.~\eqref{eqn: proposed edl loss}\footnote{The values were manually selected from a small number of candidates.}.

A dropout rate of 0.3 was applied to the transformer parameters. The system was implemented using PyTorch and the SpeechBrain toolkit~\cite{speechbrain}. The Adam optimizer was used with an initial learning rate set to 0.001. Training took $\sim$ 8 hours on an NVIDIA A100 GPU.

\begin{table*}[tb]
\centering
\scalebox{0.88}{
\begin{tabular}{l|ccc|ccc|ccc|ccc}
\toprule
  & \multicolumn{3}{c|}{\textbf{CCC $\uparrow$}} 
  & \multicolumn{3}{c|}{\textbf{RMSE $\downarrow$}}
  & \multicolumn{3}{c|}{\textbf{NLL(avg)} $\downarrow$} & \multicolumn{3}{c}{\textbf{NLL(all)} $\downarrow$} \\
% \midrule
\multicolumn{1}{c|}{\textbf{MSP-Podcast}}  & \textbf{v}    & \textbf{a}     & \textbf{d}     &
\textbf{v}    & \textbf{a}     & \textbf{d}     &\textbf{v}    & \textbf{a}     & \textbf{d}        & \textbf{v}    & \textbf{a}     & \textbf{d}       \\
\midrule
$\Loss$ in 
Eqn.~\eqref{eqn: proposed edl loss}
& {0.506}  & {0.698} & {0.613} 
& 0.772	&0.680	&0.576 
& 1.334    & 1.285   & 1.156   
& {1.696}    & {1.692}   & {1.577}  \\
$\Loss^{\sigma}=0$        
& 0.451 & 0.687 & 0.607 
& 0.784	& 0.679	& 0.580
& 1.345    & 1.277   & 1.159   
& 1.706    & 1.705   & 1.586   \\
$\Lnll =\barLnll$
& 0.473 & 0.682 & 0.609 
& 0.808	& 0.673	& 0.566
& {1.290}    & {1.060 }  & {0.899}   
& 2.027    & 2.089   & 1.969   \\
\midrule
\midrule
\multicolumn{1}{c|}{\textbf{IEMOCAP}}   
& \textbf{v}    & \textbf{a}     & \textbf{d}      &
\textbf{v}    & \textbf{a}     & \textbf{d}      &\textbf{v}    & \textbf{a}     & \textbf{d}        & \textbf{v}    & \textbf{a}     & \textbf{d}        \\
\midrule
$\Loss$ in 
Eqn.~\eqref{eqn: proposed edl loss}          
& {0.596} & 0.755 & {0.569} 
& 0.755	& 0.457	& 0.638
& 1.070 & {0.795} & {1.035} 
& {1.275} & {1.053} & {1.283}\\
$\Loss^{\sigma}=0$       
& 0.582 & 0.752 & 0.553 
& 0.772	& 0.466	& 0.655
& 1.180 & 0.773 & 1.061 
& 1.408 & 1.069 & 1.294 \\
$\Loss^{\text{NLL}}=\bar{\Loss}^{\text{NLL}}$& 0.585 & {0.759} & 0.555 
& 0.786	& 0.444	& 0.633
& {1.001} & {0.727} & 1.036 
& 1.627 & 1.329 & 1.441 \\
\bottomrule
\end{tabular}}
\caption{DEER results variations of the loss in Eqn.~\eqref{eqn: proposed edl loss}. `v' , `a', `d' stands for valence, arousal, dominance. `$\uparrow$' denotes the higher the better, `$\downarrow$' denotes the lower the better. The `$\Loss$ in 
Eqn.~\eqref{eqn: proposed edl loss}' row systems used the complete total loss of DEER. The `$\Loss^{\sigma}=0$' row systems had no $\Loss^{\sigma}$ regularisation term in the total loss. The `$\Loss^{\text{NLL}}=\bar{\Loss}^{\text{NLL}}$' row systems replaced the individual human labels with $\bar{\Loss}^{\text{NLL}}$ in the total loss. }
\label{tab: edl-loss}
\end{table*}

\subsection{Evaluation metrics}
\subsubsection{Mean prediction}
Following prior work in continuous emotion recognition~\cite{ringeval2015avec,ringeval2017avec,sridhar2020ensemble,leem2022not}, the concordance correlation coefficient (CCC) was used to evaluate the predicted mean. CCC combines
the Pearson’s correlation coefficient with the square difference between the mean of the two compared sequences:
\begin{equation*}
    \rho_{\text{ccc}}=\frac{2 \rho\,\sigma_\text{ref} \sigma_\text{hyp}}{\sigma_\text{ref}^{2}+\sigma_\text{hyp}^{2}+\left(\mu_\text{ref}-\mu_\text{hyp}\right)^{2}},
\end{equation*}
where $\rho$ is the Pearson correlation coefficient between a hypothesis sequence (system predictions) and a reference sequence, where $\mu_\text{hyp}$ and $\mu_\text{ref}$ are the mean values, and $\sigma_\text{hyp}^{2}$ and $\sigma_\text{ref}^{2}$ are the variance values of the two sequences. Hypotheses that are well correlated with the reference but shifted in value are penalised in proportion to the deviation. The value of CCC ranges from -1 (perfect disagreement) to 1 (perfect agreement).

The root mean square error (RMSE) averaged over the test set is also reported. Since the average of the human labels, $\bar{y}$, is defined as the ground truth in both datasets, $\bar{y}$ were used as the reference in computing the CCC and RMSE. However, using $\bar{y}$ also indicates that these metrics are less informative when the aleatoric uncertainty is large.

\subsubsection{Uncertainty estimation}
It is common to use NLL to measure the uncertainty estimation ability~\citep{gal2016dropout,amini2020deep}. NLL is computed by fitting data to the predictive posterior $\mathrm{q}(y)$.

In this paper, NLL(avg) defined as $-\log\mathrm{q}(\bar{y})$ and NLL(all) defined as
$-\frac{1}{M}\sum_{m=1}^M \log \mathrm{q}(y^{(m)})$ are both used. 
NLL(avg) measures how much the averaged label $\bar{y}$ fits into the predicted posterior distribution, and NLL(all) measures how much every single human label $y^{(m)}$ fits into the predicted posterior. A lower NLL indicates better uncertainty estimation.

\section{Experiments and Results}
\label{sec: exp}
\subsection{Effect of the aleatoric regulariser $\Loss^{\sigma}$}
First, by setting $\Loss^{\sigma}=0$ in the total loss, an ablation study of the effect of the proposed extra regularising term $\Loss^{\sigma}$ is performed. The results are given in the `$\Loss^{\sigma}=0$' rows in Table~\ref{tab: edl-loss}. In this case, only $\Loss^{\mu}$ is used to regularise $\Lnll$ and the results are compared to those trained using the complete loss defined in Eqn.~\eqref{eqn: proposed edl loss}, which are shown in the `$\Loss$ in 
Eqn.~\eqref{eqn: proposed edl loss}' rows.
From the results, $\Loss^\sigma$ improves the performance in CCC and NLL(all), but not in NLL(avg), as expected.

\subsection{Effect of the per-observation-based $\Lnll$}
Next, the effect of our proposed per-observation-based NLL loss defined in Eqn.~\eqref{eqn: nll-per-sample}, $\Lnll$, is compared to an alternative. Instead of using $\Lnll$,
% \vspace{-0.5ex}
\begin{align*}
    \barLnll=-\log \p(\bar{y}|\mathbf{\Omega})
\end{align*}
is used to compute the total loss during training, and the results are given in the `$\Loss^{\text{NLL}}=\bar{\Loss}^{\text{NLL}}$' rows in Table~\ref{tab: edl-loss}. While $\Lnll$ considers the likelihood of fitting each individual observation into the predicted posterior, $\barLnll$ only considers the averaged observation. Therefore, it is expected that using $\barLnll$ instead of $\Lnll$ yields a smaller NLL(avg) but larger NLL(all), which have been validated by the results in the table.

\begin{table*}[htb]
\centering
\scalebox{0.88}{
\begin{tabular}{c|ccc|ccc|ccc|ccc}
\toprule
\multicolumn{1}{l|}{} & \multicolumn{3}{c|}{\textbf{CCC}   $\uparrow$} 
& \multicolumn{3}{c|}{\textbf{RMSE $\downarrow$}} &
\multicolumn{3}{c|}{\textbf{NLL(avg)}  $\downarrow$} & \multicolumn{3}{c}{\textbf{NLL(all)}  $\downarrow$} \\
\multicolumn{1}{c|}{\textbf{MSP-Podcast}} & \textbf{v}    & \textbf{a}     & \textbf{d}    
& \textbf{v}    & \textbf{a}     & \textbf{d}  
& \textbf{v}    & \textbf{a}     & \textbf{d}           & \textbf{v}    & \textbf{a}     & \textbf{d}           \\
\midrule
DEER                  
& 0.506      & \textbf{0.698}      & \textbf{0.613}
& \textbf{0.772}	&0.680	&{0.576}
& \textbf{1.334}  & \textbf{1.285}   & {1.156}      
& \textbf{1.696}        & \textbf{1.692}        & \textbf{1.577}        \\
GP                   
& 0.342      & 0.595      & 0.486      
& 0.811	& \textbf{0.673}	& \textbf{0.566}
& 1.447        & 1.408        & 1.297        
& 1.727	& 1.808	& 1.592       \\
MCdp           
& 0.476      & 0.667      & 0.594      
& 0.874	& 0.702	& 0.623
& 1.680	& 1.300	& \textbf{1.071}
& 2.050	&   2.027	&   1.776     \\
Ensemble             
& \textbf{0.511}      & 0.679      & 0.608 
& 0.855	& 0.692	& 0.615
& 1.864	&   1.384	&   1.112
& 2.096	&   2.066	&   1.795       \\

\midrule
\midrule
\multicolumn{1}{c|}{\textbf{IEMOCAP}} 
& \textbf{v}    & \textbf{a}     & \textbf{d}  
& \textbf{v}    & \textbf{a}     & \textbf{d}  
& \textbf{v}    & \textbf{a}     & \textbf{d}          & \textbf{v}    & \textbf{a}     & \textbf{d}          \\
\midrule
DEER        
& \textbf{0.596} & \textbf{0.756} & \textbf{0.569} 
& \textbf{0.755}	& \textbf{0.457}	& \textbf{0.638}
&\textbf{1.070} & 0.795 & \textbf{1.035} & \textbf{1.275} & \textbf{1.053} & \textbf{1.283} \\
GP         & 0.535 & 0.717 & 0.512 
& 0.763 & 0.479 & 0.657
& 1.209 & \textbf{0.791} & 1.047 
& {1.295} & {1.205} & {1.380}\\
MCdp & 0.539 & 0.724 & 0.568 
& 0.786	& 0.561	& 0.702
& 1.291	& 0.849	& 1.133
& 1.549	& 1.325	& 1.747
 \\
Ensemble   & 0.580 & 0.754 & {0.560} 
& 0.778	& 0.476	& 0.686
& 1.296	& 0.864	& 1.110
& 1.584	& 1.218	& 1.749
 \\
\bottomrule
\end{tabular}}
\caption{Comparison with the baselines. `v', `a', `d' stands for valence, arousal, dominance. `$\uparrow$' denotes the higher the better, `$\downarrow$' denotes the lower the better. Best results in each column shown in bold.}
% \vspace{-1ex}
\label{tab: baseline compare}
\end{table*}

\subsection{Baseline comparisons}
Three baseline systems were built:
\begin{itemize}
    \item A Gaussian Process (GP) with a radial basis function kernel, trained by maximising the per-observation-based marginal likelihood.
    \item A Monte Carlo dropout (MCdp) system with a dropout rate of 0.4. During inference, the system was forwarded 50 times with different dropout random seeds to obtain 50 samples.
    \item An ensemble of 10 systems initialised and trained with 10 different random seeds.
\end{itemize}
The MCdp and ensemble baselines used the same model structure as the DEER system, except that the evidential output layer was replaced by a standard fully-connected output layer with three output units to predict the values of valence, arousal and dominance respectively. Following prior work~\cite{albadawy2018joint,atmaja2020multitask,sridhar2020modeling}, the CCC loss,
\begin{equation*}
  \Loss_\text{ccc} = 1-\rho_\text{ccc}  
\end{equation*}
 was used for training the MCdp and ensemble baselines. The CCC loss was computed based on the sequence within each mini-batch of training data. The CCC loss has been shown by previous studies to improve the continuous emotion predictions compared to the RMSE loss~\cite{povolny2016multimodal,trigeorgis2016adieu,le2017discretized}. 
For MCdp and ensemble, the predicted distribution of the emotion attributes were estimated based on the obtained samples by kernel density estimation. 

The results are listed in Table~\ref{tab: baseline compare}. The proposed DEER system outperforms the baselines on most of the attributes and the overall values. In particular, DEER outperforms all baselines consistently in the NLL(all) metric.

\subsection{Cross comparison of mean prediction}
Table~\ref{tab: cross compare} compares results obtained with those previously published in terms of the CCC value. Previous papers have reported results on both version 1.6 and 1.8 of the MSP-Podcast dataset. For comparison, we also conducted experiments on version 1.6 for comparison. Version 1.6 of MSP-Podcast database is a subset of version 1.8 and contains 34,280 segments for training, 5,958 segments for validation and 10,124 segments for testing.
For IEMOCAP, apart from training on Session 1-4 and testing on Session 5 (Ses05), we also evaluated the proposed system by a 5-fold cross-validation (5CV) based on a ``leave-one-session-out'' strategy. In each fold, one session was left out for testing and the others were used for training. The configuration is speaker-exclusive for both settings.
As shown in Table~\ref{tab: cross compare}, our  DEER systems achieved state-of-the-art results on both versions of MSP-Podcast and both test settings of IEMOCAP.

\begin{table*}[tb]
    \centering
    \scalebox{0.88}{
    \begin{tabular}{c|c|cccc|c}
    \toprule
    \multirow{8}{*}{\textbf{MSP-podcast}} &\textbf{Paper} & \textbf{Version} & \textbf{v} & \textbf{a} & \textbf{d} & \textbf{Average}\\
    \cmidrule{2-7}
    &\citet{ghriss2022sentiment} & 1.6 & 0.412 & 0.679 & 0.564 & 0.552 \\
    &\citet{mitra22_interspeech} & 1.6 & 0.57 & 0.75 & 0.67 & 0.663\\
    &\citet{srinivasan2022representation} & 1.6 &  0.627  &  0.757  & 0.671 & 0.685\\
    & DEER & 1.6 & \textbf{0.629} &	\textbf{0.777} & \textbf{0.684} & \textbf{0.697}\\
    \cmidrule{2-7}
    &\citet{leem2022not} & 1.8 & 0.212 & 0.572 & 0.505 & 0.430\\
    & DEER & 1.8 & \textbf{0.506} &    \textbf{0.698} &    \textbf{0.613} & \textbf{0.606}\\
\midrule
\midrule
    \multirow{7}{*}{\textbf{IEMOCAP}} & \textbf{Paper} & \textbf{Setting} & \textbf{v} & \textbf{a} & \textbf{d} & \textbf{Average} \\
    \cmidrule{2-7}
    &\citet{atmaja2020improving} &  Ses05   &   0.421   &   0.590   &   0.484   &   0.498\\
    &\citet{atmaja2021two}    &  Ses05    & 0.553 &   0.579   &   0.465   &   0.532\\
    & DEER & Ses05 &\textbf{0.596}	&\textbf{0.756}	&\textbf{0.569} & \textbf{0.640}\\
    \cmidrule{2-7}
    &\citet{srinivasan2022representation}   &   5CV  &   0.582  &   {0.667}   &   {0.545}   & 0.598\\
    & DEER &  5CV & \textbf{0.625} & \textbf{0.720} & \textbf{0.548} & \textbf{0.631} \\
    \bottomrule
    \end{tabular}}
    \caption{Cross comparison of the CCC value on MSP-Podcast and IEMOCAP. `v', `a', `d' stands for valence, arousal, dominance. `Version' of MSP-Podcast denotes the release version of the  dataset, and only the results from the same dateset version are comparable. `Test set' of IEMOCAP denotes the train/test split. `Ses05' denotes training on Session 1-4 and testing on Session 5. `5CV' denotes leave-one-session-out 5-fold cross validation.}
    \label{tab: cross compare}
\end{table*}

\begin{figure}[t]
    \centering
    \begin{minipage}[b]{\linewidth}
    \centerline{\includegraphics[width=\linewidth]{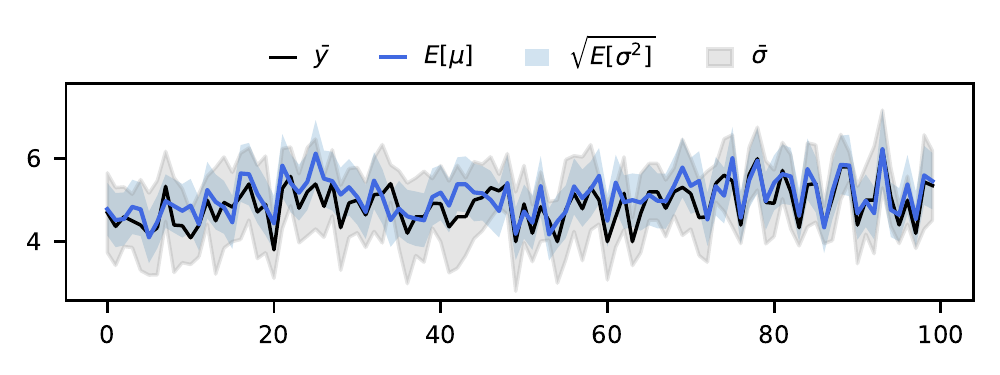}}
 \vspace{-1.5ex}
  \centerline{(a) Aleatoric uncertainty}
    \end{minipage}
    \begin{minipage}[b]{\linewidth}
    \centerline{\includegraphics[width=\linewidth]{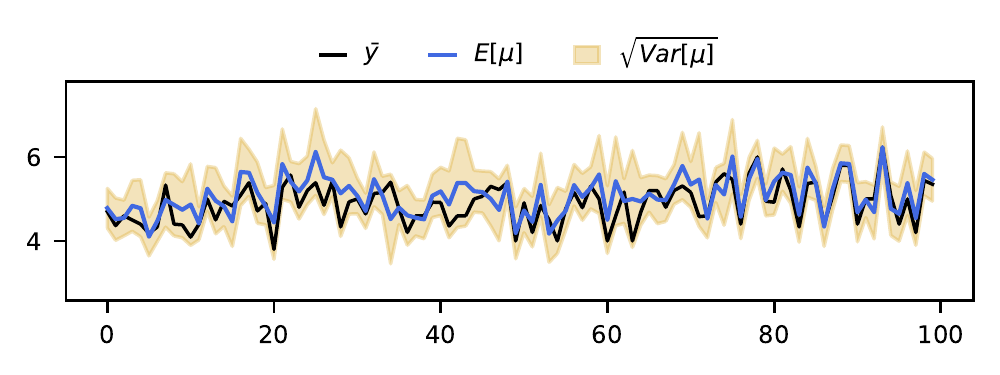}}
 \vspace{-1.5ex}
  \centerline{(b) Epistemic uncertainty}
    \end{minipage}
    \begin{minipage}[b]{\linewidth}
    \centerline{\includegraphics[width=\linewidth]{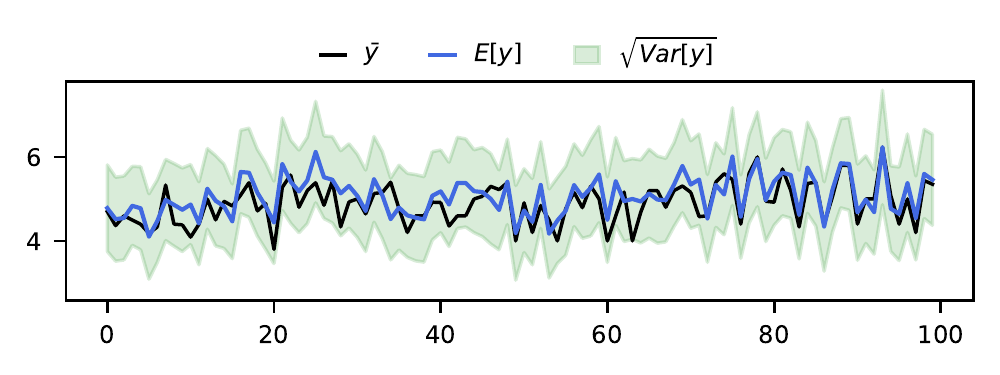}}
 \vspace{-1.5ex}
  \centerline{(c) Total uncertainty}
    \end{minipage}
    \caption{Visualisation of (a) aleatoric (b) epistemic (c) total uncertainty of dominance for MSP-Podcast. $x$-asix is the test utterance index.}
    \label{fig: plot}
    \vspace{-2ex}
\end{figure}

\subsection{Analysis of uncertainty estimation}
\subsubsection{Visualisation}

Based on a randomly selected subset test set of MSP-Podcast version 1.8, the aleatoric, epistemic and total uncertainty of the dominance attribute predicted by our proposed DEER system are shown in Figure~\ref{fig: plot}.

Figure~\ref{fig: plot} (a) shows the predicted mean  $\pm$ square root of the predicted aleatoric uncertainty ($\E[\mu] \pm \sqrt{\E[\sigma^2]}$) and the average label $\pm$ the standard deviation of the human labels ($\bar{y} \pm \bar{\sigma}$). It can be seen that the predicted aleatoric uncertainty (blue) overlaps with the label standard deviation (grey) and the overlapping is more evident when the mean predictions are accurate ( {i.e.} samples around index 80-100).

Figure~\ref{fig: plot} (b) shows the predicted mean $\pm$ square root of the predicted epistemic uncertainty ($\E[\mu] \pm \sqrt{\Var[\mu]}$). The epistemic uncertainty is high when the predicted mean deviates from the target ( {i.e. }samples around index 40-50) while low then the predicted mean matches the target ( {i.e. }samples around index 80-100). 

Figure~\ref{fig: plot} (c) shows the predicted mean $\pm$ square root of the total epistemic uncertainty ($\E[y] \pm \sqrt{\Var[y]}$) which combines the aleatoric and epistemic uncertainty. The total uncertainty is high either when the input utterance is complex or the model is not confident.

\subsubsection{Reject option}
A reject option was applied to analyse the uncertainty estimation performance, where the system has the option to accept or decline a test sample based on the uncertainty prediction. 
Since the evaluation of CCC is based on the whole sequence rather than individual samples, its computation would be affected when the sequence is modified by rejection~\cite{wu2022novel}.
Therefore, the reject option is performed based on RMSE. 

\begin{figure}[tb]
    \centering
    \begin{minipage}[b]{0.49\linewidth}
\centerline{\includegraphics[width=\linewidth]{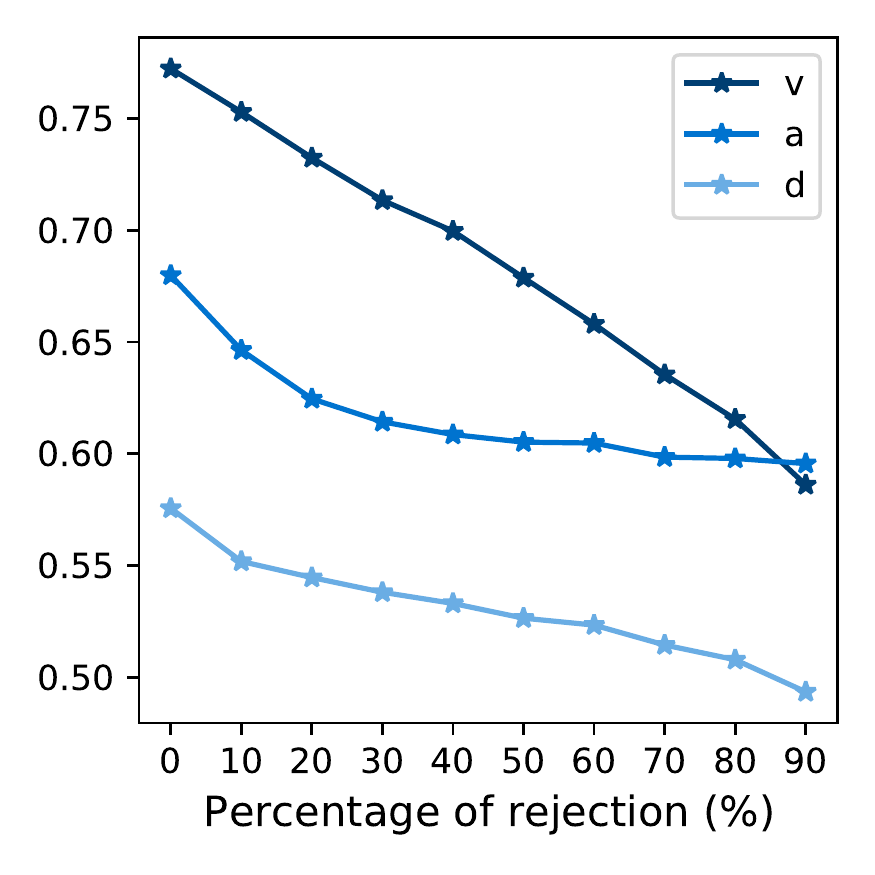}}
  \centerline{(a) MSP-Podcast}
    \end{minipage}
    \begin{minipage}[b]{0.49\linewidth}
    \centerline{\includegraphics[width=\linewidth]{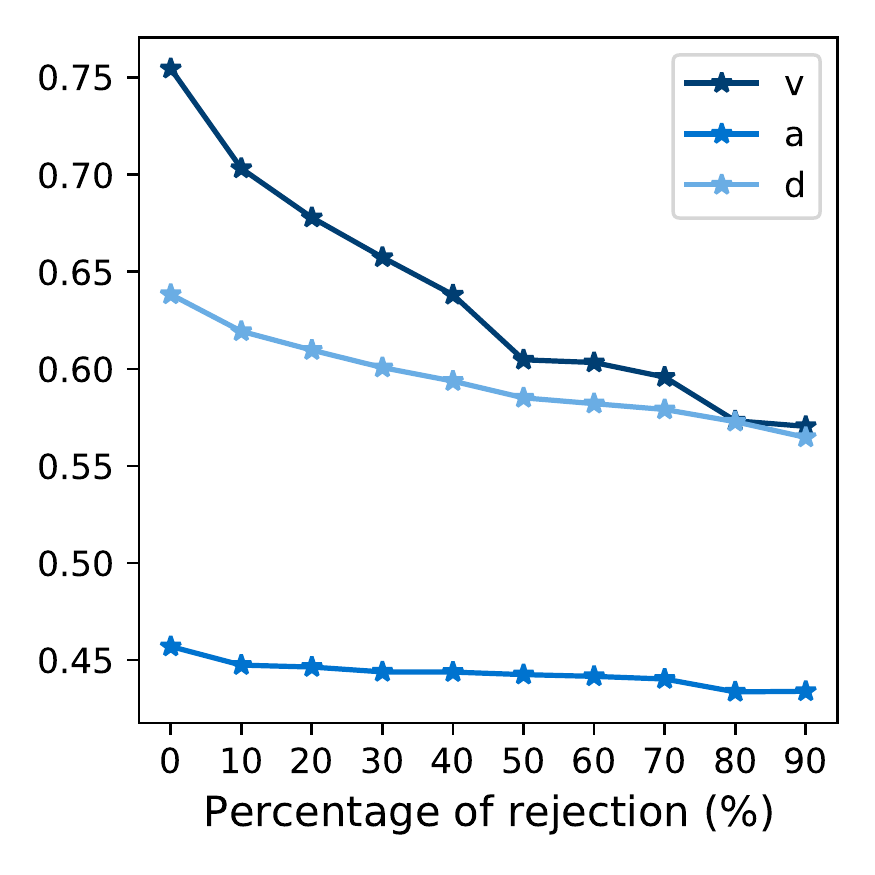}}
  \centerline{(b) IEMOCAP}
    \end{minipage}
    \caption{Reject Option of RMSE based on predicted variance for (a) MSP-Podcast and (b) IEMOCAP.}
    \label{fig: rej}
\end{figure}

Confidence is measured by the total uncertainty given in Eqn.~\eqref{eqn: predicted mean var}. Figure~\ref{fig: rej} shows the performance of the proposed DEER system with a reject option on MSP-Podcast and IEMOCAP. A percentage of  utterances with the largest predicted variance were 
rejected. The results at 0\% rejection corresponds to the RMSE achieved on the entire test data. As the percentage of rejection increases, test coverage decreases and the average RMSE decreases showing the predicted variance succeeded in confidence estimation. The system then trades off between the test coverage and performance.

\FloatBarrier
\section{Conclusions}
Two types of uncertainty exist in AER: (i) aleatoric uncertainty arising from the inherent ambiguity of emotion and personal variations in emotion expression; (ii) epistemic uncertainty associated with the estimated network parameters given the observed data. This paper proposes DEER for estimating those uncertainties in emotion attributes. Treating observed attribute-based annotations as samples drawn from a Gaussian distribution, DEER places a normal-inverse gamma (NIG) prior over the Gaussian likelihood.  A novel training loss is proposed which combines a per-observation-based NLL loss with a regulariser on both the mean and the variance of the Gaussian likelihood. Experiments on the MSP-Podcast and IEMOCAP datasets show that DEER can produce  state-of-the-art results in estimating both the mean value and the distribution of emotion attributes. The use of NIG, the conjugate prior to the Gaussian distribution, leads to tractable analytic computation of the marginal likelihood as well as aleatoric and epistemic uncertainty associated with attribute prediction. Uncertainty estimation is analysed by visualisation and a reject option. Beyond the scope of AER, DEER could also be applied to other tasks with subjective evaluations yielding inconsistent labels. 

\section*{Limitations}
The proposed approach (along with other methods for estimating uncertainty in inconsistent annotations) is only viable when the raw labels from different human annotators for each sentence are provided by the datasets. However, some multiple-annotated datasets only released the majority vote or averaged label for each sentence ( {i.e.} \citealp{Meld}).

The proposed method made a Gaussian assumption on the likelihood function for the analytic computation of the uncertainties. The results show that this modelling approach is effective. Despite the effectiveness of the proposed method, other distributions could also be considered.

Data collection processes for AER datasets vary in terms of recording conditions, emotional elicitation scheme, and annotation procedure,  {etc}. This work was tested on two typical datasets: IEMOCAP and MSP-Podcast. The two datasets are both publicly available and differ in various aspects: 
\vspace{-1ex}
\begin{itemize}
    \item IEMOCAP contains emotion acted by professional actors while MSP-Podcast contains natural emotion.
    \vspace{-1ex}
    \item  IEMOCAP contains dyadic conversations while MSP-Podcast contains Podcast recordings.
    \vspace{-1ex}
    \item IEMOCAP contains 10 speakers and MSP-Podcast contains 1285 speakers.
    \vspace{-1ex}
    \item IEMOCAP contains about 12 hours of speech and MSP-Podcast contains more than 110 hours of speech.
    \vspace{-1ex}
    \item IEMOCAP was annotated by six professional evaluators with each sentence being annotated by three evaluators. MSP-Podcast was annotated by crowd-sourcing where a total of 11,799 workers were involved and each work annotated 41.5 sentences on average.
\end{itemize}
The proposed approach has been shown effective over both datasets. We believe the proposed technique should be generic. Furthermore, although validated only for AER, the proposed method could also be applied to other tasks with disagreements in subjective annotations such as hate speech detection and language assessment.

\section*{Ethics Statement}
In tasks involving subjective evaluations such as emotion recognition, it is common to employ multiple human annotators to give multiple annotations to each data instance. When annotators disagree, majority voting and averaging are commonly used to derive single ground truth labels for training supervised machine learning systems. However, in many subjective tasks, there is usually no single ``correct'' answer. By enforcing a single ground truth, there's a potential risk of ignoring the valuable nuance in each annotator's evaluation and their disagreements. This can cause minority views to be under-represented. The DEER approach proposed in this work could be beneficial to this concern as it models uncertainty in annotator disagreements and provides some explainability of the predictions.

While our method helps preserve minority perspectives, misuse of this technique might lead to ethical concerns. Emotion recognition is at risk of exposing a person's inner state to others and this information could be abused. Furthermore, since the proposed approach takes each annotation into consideration, it is important to protect the anonymity of annotators.

\section*{Acknowledgements}
Wen Wu is supported by a Cambridge International Scholarship from the Cambridge Trust. This work has been performed using resources provided by the Cambridge Tier-2 system operated by the University of Cambridge Research Computing Service (www.hpc.cam.ac.uk) funded by EPSRC Tier-2 capital grant EP/T022159/1.

The MSP-Podcast data was provided by The University of Texas at Dallas through the Multimodal Signal Processing Lab. This material is based upon work supported by the National Science Foundation under Grants No. IIS-1453781 and CNS-1823166. Any opinions, findings, and conclusions or recommendations expressed in this material are those of the author(s) and do not necessarily reflect the views of the National Science Foundation or The University of Texas at Dallas.

\bibliography{acl}

\appendix

\section{Derivation of the predictive posterior}
\label{sec:appendix}

Since NIG is the Gaussian conjugate prior,
\begin{equation}
 \begin{aligned}
    \nonumber \p(\boldsymbol{\Psi}|\boldsymbol{\Omega})
    &= \mathcal{N}(\gamma, \sigma^2 \upsilon^{-1}) \, \Gamma^{-1}(\alpha, \beta)\\
    &= \frac{\beta^{\alpha}\sqrt{\upsilon}}{\Gamma(\alpha)\sqrt{2 \pi \sigma^{2}}} \left(\frac{1}{\sigma^{2}}\right)^{\alpha+1}\\
    &\cdot \exp \left\{-\frac{2 \beta+\upsilon(\gamma-\mu)^{2}}{2 \sigma^{2}}\right\}
\end{aligned}   
\end{equation}
its posterior $\p(\boldsymbol{\Psi}|\mathcal{D})$ is in the same parametric family as the prior $\p(\boldsymbol{\Psi}|\boldsymbol{\Omega})$. Therefore, given a test utterance $\x_*$, the predictive posterior $\p(y_*|\mathcal{D})$ has the same form as the marginal likelihood $\p(y|\boldsymbol{\Omega})$, where $\mathcal{D}$ denotes the training set.
\begin{align}
\p(y_*|\mathcal{D}) = \int \p(y_*|\boldsymbol{\Psi}) \p(\boldsymbol{\Psi} |\mathcal{D}) \diff \boldsymbol{\Psi}\\
    \p(y|\boldsymbol{\Omega}) = \int \p(y|\boldsymbol{\Psi}) \p(\boldsymbol{\Psi} |\boldsymbol{\Omega}) \diff \boldsymbol{\Psi}
\end{align}

In DEER, the predictive posterior and posterior are both conditioned on $\boldsymbol{\Omega}$, written as $\p(y_*|\mathcal{D},\boldsymbol{\Omega})$ and $\p(\boldsymbol{\Psi}|\mathcal{D},\boldsymbol{\Omega})$ to be precise. Also, the information of $\mathcal{D}$ is contained in $\boldsymbol{\Omega}_*$ since $\boldsymbol{\Omega}_* = f_{ \hat{\bm \Theta}}(\x_*)$ and $\hat{\bm \Theta}$ is the optimal model parameters obtained by training on $\mathcal{D}$. Then the predictive posterior can be written as $\p(y_*|\boldsymbol{\Omega}_*)$. Given the conjugate prior, the predictive posterior in DEER can be computed by directly substituting the predicted $\boldsymbol{\Omega}_*$ into the expression of marginal likelihood derived in Eqn.~\eqref{eq:post_pred}, skipping the step of calculating the posterior.

\section{Fusion with text modality}
\label{sec:fuse}
\begin{figure}[tb]
    \centering
    \includegraphics[width=\linewidth]{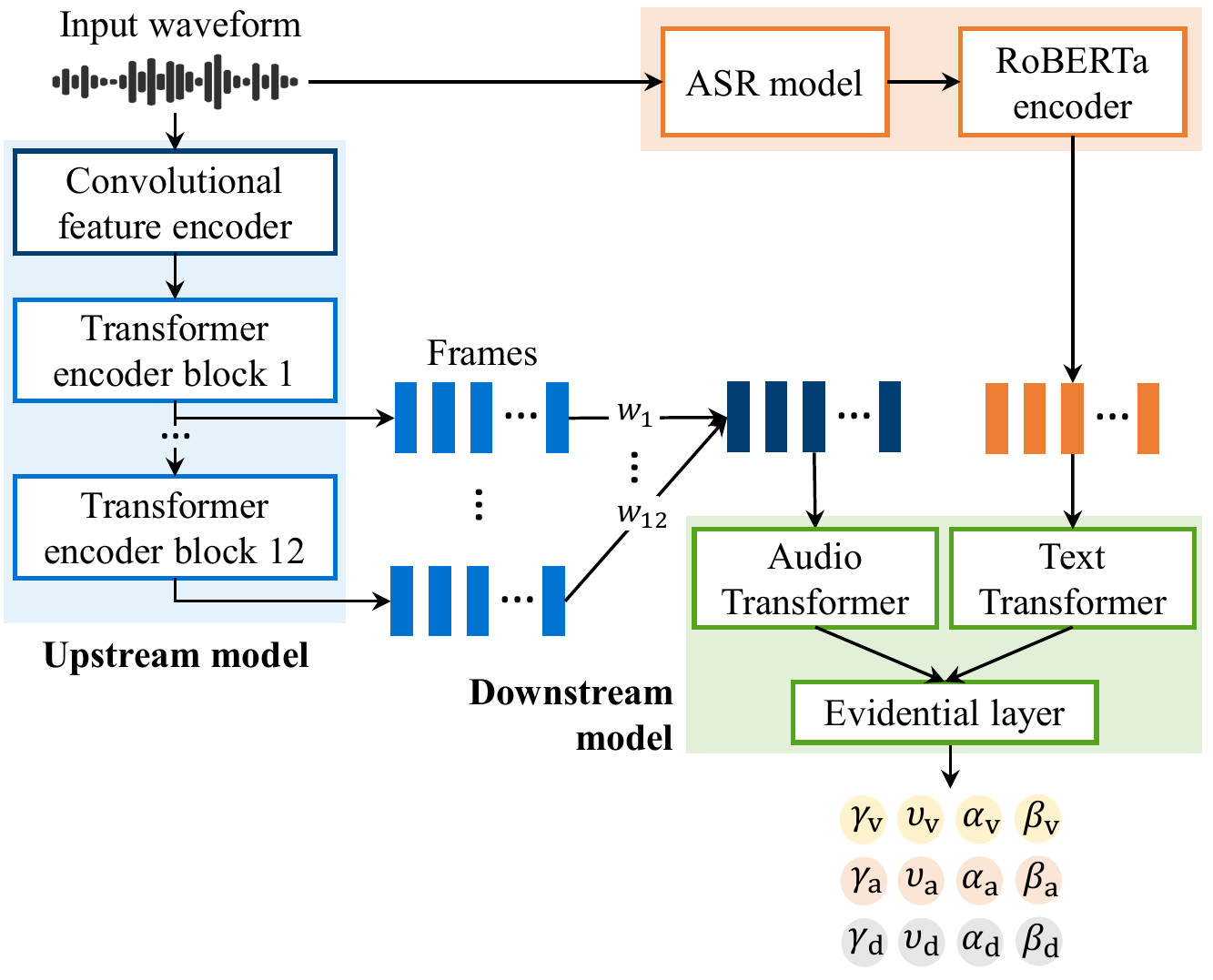}
    \caption{Model structure for bi-modal experiments. }
    \label{fig: struc-fuse}
\end{figure}

This appendix presents bi-modal experiments that incorporate text information into the DEER model. Transcriptions were obtained from a publicly available automatic speech recognition (ASR) model ``wav2vec2-base-960h" \footnote{https://huggingface.co/facebook/wav2vec2-base-960h} which fine-tuned the wav2vec 2.0~\cite{baevski2020wav2vec} model on 960 hours Librispeech data~\cite{panayotov2015librispeech}. 
Transcriptions were first encoded by a RoBERTa model~\cite{liu2019roberta} and fed into another two-layer Transformer encoder. As shown in Figure~\ref{fig: struc-fuse}, outputs from the text Transformer were concatenated with the outputs from the audio Transformer encoder and fed into the evidential output layer. 
\begin{table}[htb]
    \centering
    \scalebox{0.88}{
    \begin{tabular}{c|c|ccc}
    \toprule
    \multirow{3}{*}{\textbf{MSP-podcast}} &\textbf{Modality} &  \textbf{v} & \textbf{a} & \textbf{d}\\
    \cmidrule{2-5}
    & A &{0.506} &    {0.698} &    {0.613} \\
    & A+T & 0.559 & 0.699 & 0.614\\
\midrule
\midrule
    \multirow{3}{*}{\textbf{IEMOCAP}} & \textbf{Modality}  & \textbf{v} & \textbf{a} & \textbf{d} \\
    \cmidrule{2-5}
    & A &{0.596}	&{0.756}	&{0.569} \\
    & A+T  & {0.609}	&{0.754}	&{0.575} \\
    \bottomrule
    \end{tabular}}
    \caption{CCC value for bi-modal experiments. `A' and `T' stands for audio and text. `v', `a', and `d' stand for valence, arousal, and dominance. Release 1.8 is used for MSP-Podcast. `Ses05' setup used for IEMOCAP that trains on Session 1-4 and tests on Session 5.}
    \label{tab: fuse}
\end{table}
Results are shown in Table~\ref{tab: fuse}. Incorporating text information improves the estimation of valence but not necessarily for arousal and dominance. Similar phenomena were observed by~\cite{triantafyllopoulos22b_interspeech}. A possible explanation is that text is effective for sentiment analysis (positive or negative) but may not be as informative as audio to determine a speaker's level of excitement. CCC for dominance improves more for IEMOCAP than MSP-Podcast possibly because IEMOCAP is an acted dataset and the emotion may be exaggerated compared with MSP-Podcast which contains natural emotion.

\end{document}